\documentclass[aps,pra,superscriptaddress, twocolumn, amsfonts, amsmath, amssymb, floatfix]{revtex4-2}
\usepackage{graphicx}
\usepackage{float}
\usepackage{sidecap}
\usepackage{dcolumn}
\usepackage{bm}
\usepackage{epsfig}
\usepackage{braket}
\usepackage{color}
\usepackage{ulem}
\usepackage{amsmath}
\usepackage[colorlinks=true, pdfstartview=FitV, linkcolor=blue, citecolor=blue, urlcolor=blue]{hyperref}

\begin{document}

\title{Floquet-engineered moiré quasicrystal patterns of ultracold Bose gases in twisted bilayer optical lattices}


\author{Zhenze Fan}
\affiliation{Department of Physics, School of Physics, East China Normal University, Shanghai 200241, China}

\author{Juan Wang}
\affiliation{Department of Physics, School of Physics, East China Normal University, Shanghai 200241, China}

\author{Yan Li}
\email{yli@phy.ecnu.edu.cn}
\affiliation{Department of Physics, School of Physics, East China Normal University, Shanghai 200241, China}
\affiliation{Chongqing Key Laboratory of Precision Optics, Chongqing Institute of East China Normal University, Chongqing 401120, China}

\begin{abstract}
We investigate the formation of moiré quasicrystal patterns in Bose gasses confined in twisted bilayer optical lattices via Floquet-engineered intralayer atomic interactions. Dynamical evolutions of the total density wave amplitude exhibit the stage for the emergence of moiré quasicrystal patterns, where the pattern formation is closely associated with the momenta of collective modes excited by the weak periodic drive. Through analyzing the radial and angular density wave amplitude, we find that these new collective modes are only coupled radially and cannot be decoupled eventually. The symmetry of quasicrystal patterns can be easily manipulated by the modulation frequencies and amplitudes.  Reducing the frequencies and increasing the amplitudes can both facilitate lattice symmetry breaking and the subsequent emergence of rotational symmetry. Notably, a twelve-fold quasicrystal pattern emerges under specific parameters, closely resembling the moiré quasicrystal in twisted bilayer graphene. The momentum-space distributions also exhibit high rotational symmetry, which is consistent with the real-space patterns at specific evolution times. Our findings establish a new quantum platform for exploring quasicrystals and their symmetry properties in ultracold bosonic systems.
\end{abstract}

\maketitle


\section{Introduction}
The recent appearance of twisted two-dimensional (2D) materials has profoundly impacted condensed-matter physics. In particular, the discovery of unconventional superconductivity \cite{Unconventionalsuperconductivity} and correlated insulating behavior \cite{Correlatedinsulator} in twisted bilayer graphene has revolutionized the field of moiré physics. A moiré lattice refers to a new periodic lattice structure formed when two identical periodic lattices are rotated by a specific angle. Moiré lattices with a small twist angle host rich physical properties, such as flat bands \cite{flat2019,flat2021,flat2021.2}, quantum anomalous Hall effect \cite{Hall2020,Hall2023,Hall2023.2,Hall2023.3,Hall2023.4,Hall2024}, moiré excitons \cite{excition2019.1,excition2019.2,excition2021,excition2023}, magnetism \cite{ferromagnetism2019.1,ferromagnetism2019.2,ferromagnetism2019.3,ferromagnetism2020,ferromagnetism2020.1,ferromagnetism2021} and strongly correlated insulators \cite{insulator2018,insulator2019.1,insulator2019.2,insulator2021}.

Ultracold Bose gases in optical lattices are becoming an ideal platform for studying twisted bilayer systems, owing to their purity and high tunability. Quantum simulations in twisted bilayer optical lattices are realized by interfering multiple sets of laser beams to create the desired lattice geometries. Different atomic hyperfine spin states are employed to construct spin-dependent lattices of two-component bosonic systems. The schemes utilizing spin-dependent square \cite{coldatom} and hexagonal \cite{hexagonal} optical lattices to simulate twisted bilayers have been proposed in the past few years. Notably, Zhang\textit{ et al.} have experimentally realized atomic Bose-Einstein condensates (BECs) in twisted bilayer lattices \cite{zhangjing}. Based on this experimental foundation, many new research findings concerning twisted bilayer lattices have been reported, including fractal structures \cite{fractal1,fractal2}, interaction-induced interlayer coupling \cite{interactioninduced,interactioninduced2}, solitons \cite{soliton1,soliton2}, topology \cite{topology1,topology2}, dipolar bosons \cite{dipolar} and quasicrystal optical lattices \cite{quasicrysta}.

Previous studies have primarily focused on the ground state of square lattices, in which quasicrystal optical lattices are obtained by adding an external quasiperiodic potential. Interestingly, when the twist angle of twisted bilayer graphene is $30^\circ$, its electronic state distributions form a moiré quasicrystal with twelve-fold ($D_{12}$) symmetries \cite{graphenequasicrystal1,graphenequasicrystal2,graphenequasicrystal3}. In bosonic systems, the competition between disorder and interaction is the key to the emergence of quasicrystal structures, which typically results in a new quantum phase named Bose glass \cite{BG1,BG2,BG3,BG}. Quasicrystal structures are controlled by externally applied quasiperiodic potentials in ultracold bosonic systems, while graphene moiré quasicrystals are primarily governed by the twist angle. 

We have investigated parametric excitations and Faraday pattern formation by periodically modulating the atomic scattering lengths in binary BECs \cite{Farady}. On the other hand, we note that Chin\textit{ et al.}  have experimentally realized hexagonal lattice density wave (DW) patterns in BECs through Floquet-engineered atomic interactions \cite{zhang2020pattern}. Meanwhile, researches on inducing novel quantum phases through Floquet engineering in optical lattices, achieved by periodically shaking the lattice or modulating atomic interactions, have been reported \cite{shaking1,shaking2,shaking3}.

This work explores the formation of novel moiré quasicrystal DW patterns in twisted bilayer hexagonal lattices via Floquet-engineered intralayer atomic interactions, in which additional quasiperiodic potentials are not necessary. In quantum systems, DW patterns represent a form of spatial order arising from nonequilibrium dynamics induced by driving fields. Investigating their formation mechanism, dynamical evolution and control is essential for understanding collective effects, symmetry breaking and restoration in complex systems.

\section{Theoretical Model}  The twisted hexagonal bilayer optical lattice of ultracold bosonic system can be realized using synthetic dimension techniques in current experiment \cite{zhangjing,spinstate,spinstate2_lamdamo}. Three laser beams of these two wavelengths intersect at $120^\circ$ in the x-y plane to produce two hexagonal lattices $V_1$ and $V_2$, where each beam is linearly polarized in-plane. A hexagonal lattice potential is given by $V(\boldsymbol{r}) = -V_0 \left| \sum_{j=1}^{3} \epsilon_j \exp[i \boldsymbol{k}_{j} \cdot (\boldsymbol{r} - \boldsymbol{r_0})] \right|^2$, with  the lattice depth $V_0$, the laser wave vector $k_j$ and the polarization $\epsilon_j$. These two spin-dependent lattices are twisted by $\theta/2$ each other, where the $\theta$ satisfies $\cos\theta = \frac{n^2 + m^2 + 4mn}{2(n^2 + m^2 + mn)}$ with two integers $(m,n)$ \cite{Theta}. The system is loaded into a harmonic trap and z-direction is tightly confined by a spin-independent potential, reducing the dynamics to quasi-2D. Meanwhile, the interlayer coupling $\Omega$ of two spin states is controlled by microwaves.

\begin{figure}
		\centering
		\includegraphics[width=1\columnwidth]{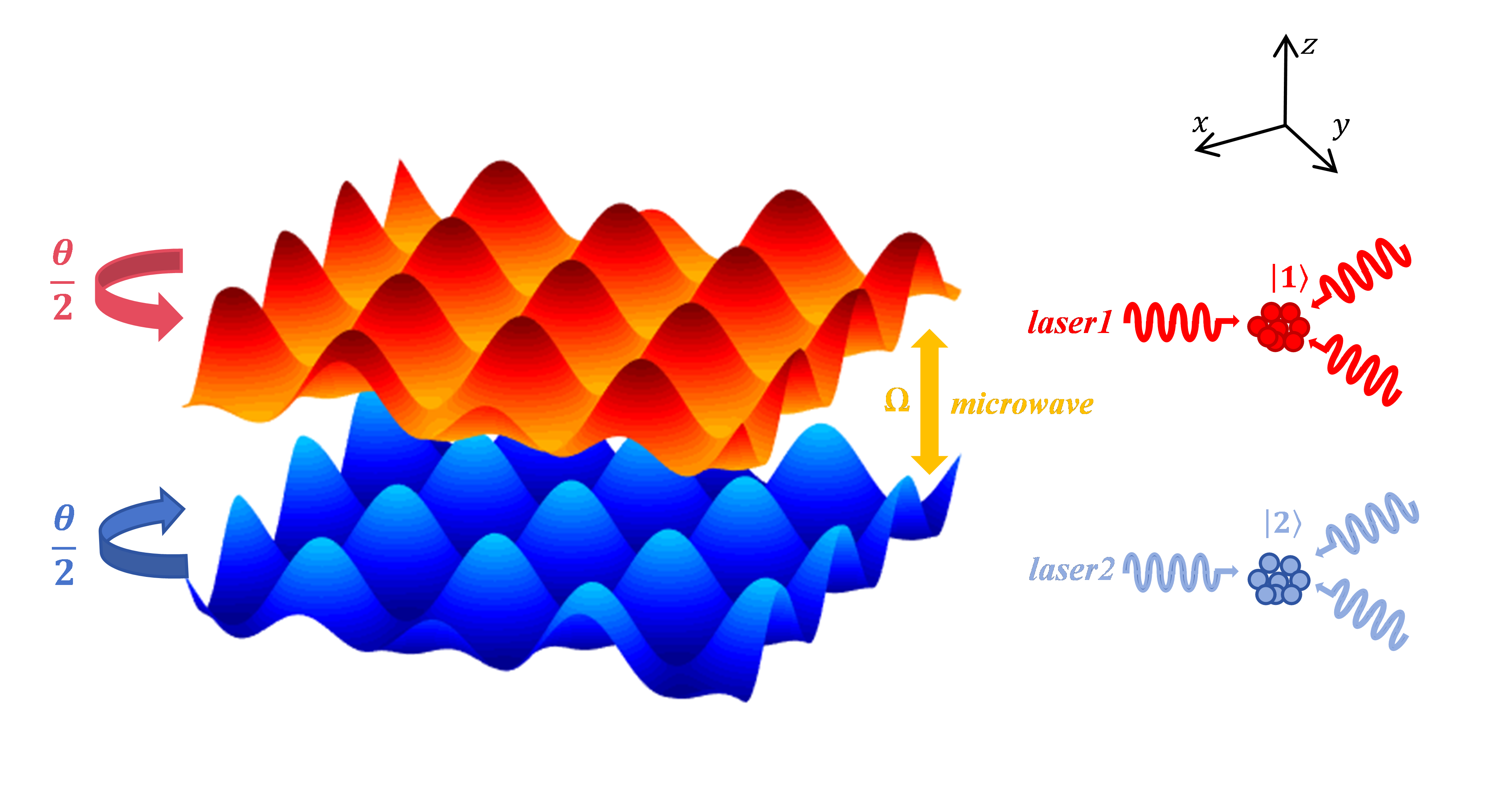}
		\caption{\fontfamily{ptm}\selectfont Cold atomic system in spin-dependent twisted bilayer hexagonal lattice. The system is loaded into a 2D harmonic trap in the x-y plane and tightly confined in the z direction, which reduces the dynamics of the system to quasi-2D.}
		\label{system1}
	\end{figure}

The ground state and dynamics of weakly interacting superfluid BECs in shallow optical lattices can be effectively studied by mean-field approximation and Gross-Pitaevskii (GP) equations. Therefore, the system's coupled GP equations reads
\begin{align}
\label{eq:GP}
i\hbar \frac{\partial \psi_1}{\partial t} = &\left( -\frac{\hbar^2}{2m} \nabla^2 + V_1+V_{trap} + g_{11} |\psi_1|^2 + g_{12} |\psi_2|^2 \right) \nonumber\\
&\psi_1+\hbar \Omega \psi_2,
\\i\hbar \frac{\partial \psi_2}{\partial t} = &\left( -\frac{\hbar^2}{2m} \nabla^2 + V_2+V_{trap} + g_{22} |\psi_2|^2 + g_{12} |\psi_1|^2 \right) \nonumber\\
&\psi_2+\hbar \Omega \psi_1,\nonumber
\end{align}
where $V_{trap}=\frac{1}{2}m(\omega_{x}^2x^2+\omega_{y}^2y^2+\omega_zz^2)$ is a harmonic trap and $\psi_i(i=1,2)$ are the wave fuctions normalized as $\int \int\left( |\psi_i|^2\right)dxdy=N$, with $N$ the total atom number. $g_{ii},g_{ij}$ characterize the intralayer and interlayer atomic interactions strength. Focus on the SU(2) symmetric interaction in calculations, we set $g_{11}=g_{22}=g_{12}=g_0$ because of the similarity in scattering lengths $a_{11}$, $a_{22}$ and $a_{12}$ for the $^{87}$Rb atoms. In our numerics, the trapping frequency is $\omega_{x,y,z} = 2\pi \times \{20, 20,1000\}$ Hz, hence the system is seen as quasi-2D because of the strong potential in z-direction. The quasi-2D interaction satisfies $g_{2D} = g_0/{\sqrt{2\pi}a_z}$, where $g_0=4\pi \hbar^2a/m$ and $a_z = \sqrt{{\hbar}/m \omega_z}$ represents the characteristic length along the z axis. Here the scattering length for the $^{87}$Rb atoms is $a=100a_0$ with $a_0$ the Bohr radius.

The GP equations is numerically solved via the imaginary time evolution method to obtain the ground state.  We make the scattering length oscillate as $a(t)=100a_0+a_m\sin{\omega t}$ by modulating the magnetic field near a Feshbach resonance \cite{Feshbach}. This paper mainly investigates the oscillating intralayer atomic interactions $g_{11}(t)=g_{22}(t)=4\pi \hbar^2a(t)/(\sqrt{2\pi}a_zm)$, while keeping the interlayer-atomic interactions $g_{12}=g_0$. Then we use the dynamical GP equations to investigate the evolution of driven DW patterns in both real and momentum spaces.

\begin{figure}
		\centering
		\includegraphics[width=1\columnwidth]{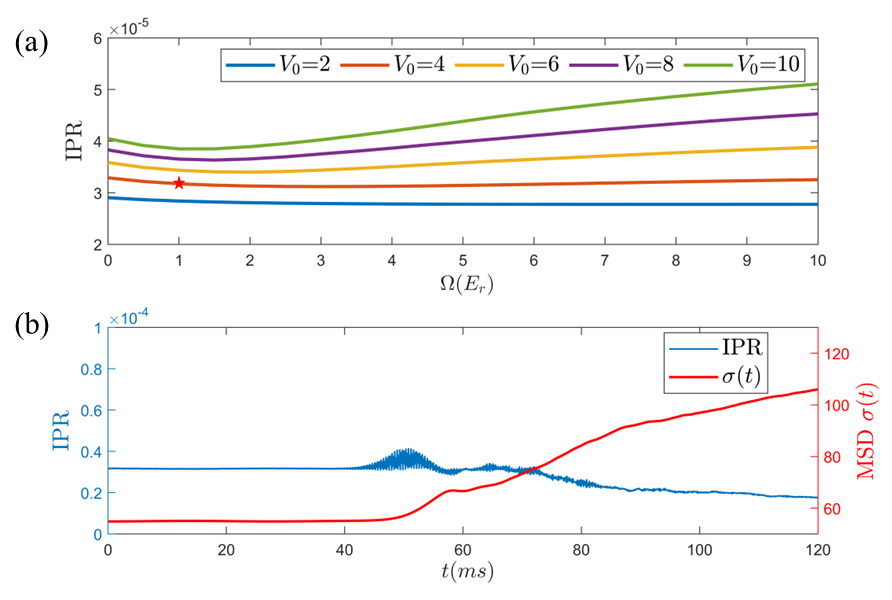}
		\caption{\fontfamily{ptm}\selectfont IPR of the system. (a) IPR of ground DW with different lattice depth $V_0$ and interlayer coupling $\Omega$, where the red star denotes the parameters employed in our following Floquet engineering. (b) IPR and MSD} of dynamical DW with $\omega=200$ Hz and $a_m=50a_0$.
		\label{IPR}
	\end{figure}

\section{Dynamical behavior}
To describe the spatial localization of the superfluid DW, we introduce a physical quantity called Inverse participation ratio (IPR) \cite{IPR1,IPR2}, which is given by 
\begin{align}
\label{eq:IPR}
\text{IPR} =\frac{\int|\psi|^{4}d\boldsymbol{r}}{(\int|\psi|^{2}d\boldsymbol{r})^2},
\end{align}
where we set $\int|\psi|^{2}d\boldsymbol{r}=1$. Figure \ref{IPR}\textcolor{blue}{(a)} illustrates the dependence of IPR on the lattice depth $V_0$ and interlayer coupling $\Omega$. The IPR increases with $V_0$, indicating a more localized DW distribution. In contrast, it exhibits a non-monotonic dependence on $\Omega$, first decreasing and then increasing. The minimum of the IPR-$\Omega$ curve is around $\Omega=E_r$ with energy unit $E_r=\hbar^2k^2/2m$, indicating the best superfluidity. In addition, the GP equations is used to describe the superfluid BECs in shallow optical lattices, but $V_0$ must remain sufficiently large to ensure a well-defined lattice distribution. Based on this condition, we select $\Omega=E_r$ and $V_0=4E_r$ as basic parameters for our  numerics (red star in Fig. \ref{IPR}\textcolor{blue}{(a)}). The other parameters used in our numerical computations are $N=2\times 10^4$ and $\theta=9.43^\circ$ (m=3, n=4).

 However, the IPR is always at the order $N^{-1}$ in the whole dynamical evolution, whose variation is too small to reflect the localization of the system. Therefore, we introduce the mean-square displacement (MSD) $\sigma(t)=[\frac{\int r^2|\psi|^{2}d\boldsymbol{r}}{\int|\psi|^{2}d\boldsymbol{r}}]^\frac{1}{2}$ \cite{interactioninduced,MSD1,MSD2} to quantitatively distinguish the different dynamical behaviors of the system at a long evolution time. $\sigma(t)$ keep a stable value at early evolution time, indicating the system remains unexcited and exhibits localized behavior. Then $\sigma(t)$ tends to linearly increase in a large evolution time, characterizing its delocalized behavior (Fig. \ref{IPR}\textcolor{blue}{(b)} red curve). Notably, this linear increase comprises different segments with distinct slopes, implying four evolution stages.
 
\section{Moiré quasicrystal patterns}

\subsection{Density wave amplitude}  
We use dynamical DW $n(\boldsymbol{r},t)=|\psi(\boldsymbol{r},t)|^{2}$ and $n(\boldsymbol{k},t)=|\psi(\boldsymbol{k},t)|^{2}$ to investigate the evolution of patterns. To characterize the stages of DW evolution, we compute the total DW amplitude, $A_{total}(\tilde{n})=\int \tilde{n}(\mathbf{k})d \mathbf{k}$ and divide the dynamical evolution into four stages by the variation of $A_{total}(\tilde{n})$. Here, $\tilde{n}(\mathbf{k})=\int e^{-i\mathbf{k}\cdot \mathbf{r}}{n}(\mathbf{r})d\mathbf{r}/2\pi$ is the  the density Fourier transformation.

\begin{figure}
		\centering
		\includegraphics[width=1\columnwidth]{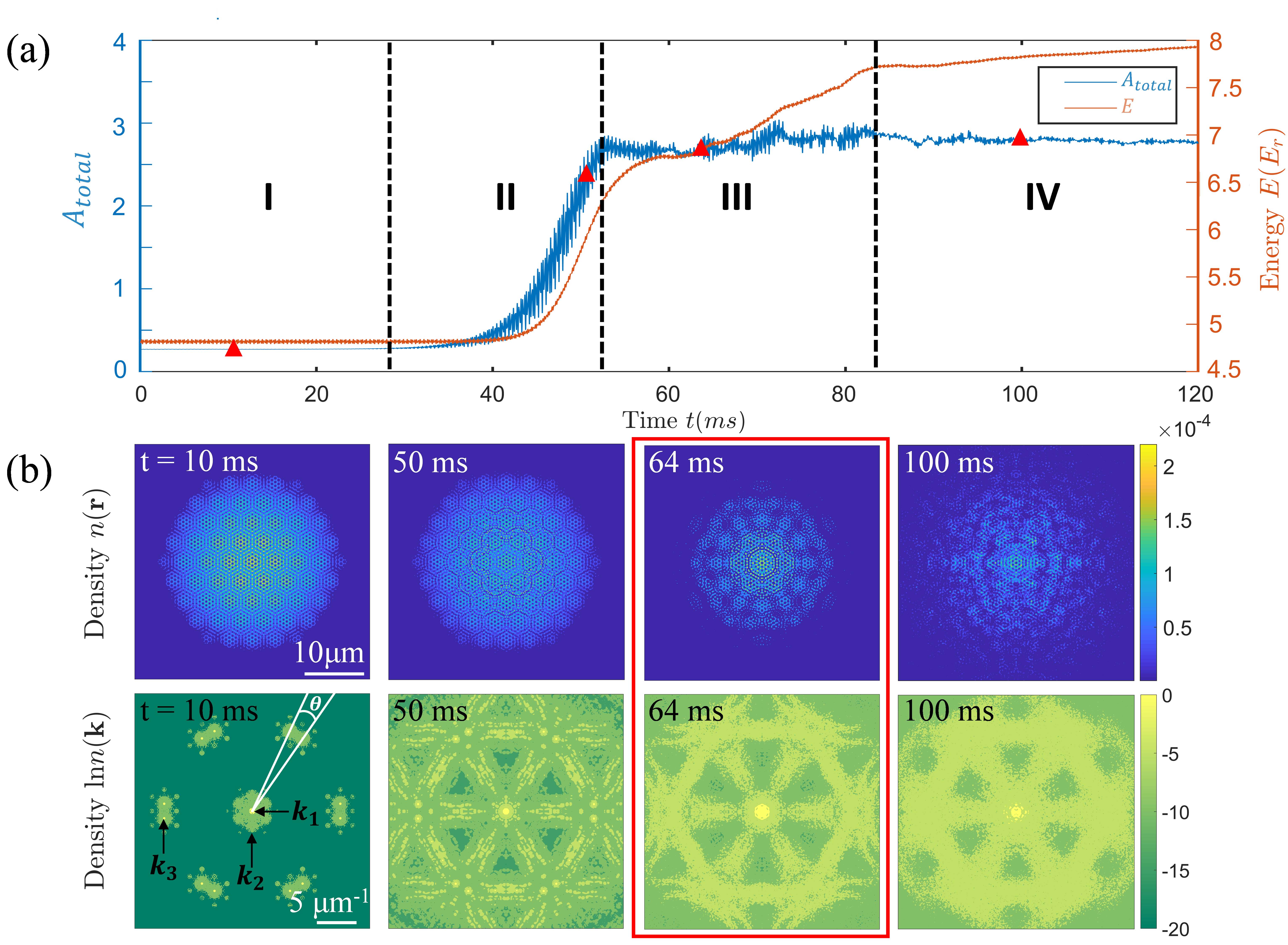}
		\caption{\fontfamily{ptm}\selectfont The total DW amplitude and patterns with $\omega=200$ Hz and $a_m=50a_0$. (a) Four evolution stages divided by $A_{total}$} (blue curve) and energy (brown curve). (b) The real-space patterns at $t=10, 50, 64, 100$ ms (top) and the corresponding momentum-space patterns amplified by logarithm (bottom). The red triangles in (a) mark the four selected temporal nodes displayed in (b).
		\label{density wave amplitude}
	\end{figure}

The evolution of driven DW patterns can be divided into four distinct stages (Fig. \ref{density wave amplitude}\textcolor{blue}{(a)}). In the preparation stage, the system remains unexcited and persistently maintains its initial moiré lattice pattern. In the excitation stage, the drive-induced excitation initiates at the central lattice and propagates outward. This process lays the foundation for subsequent lattice symmetry breaking. In the pattern-forming stage, the lattice symmetry is broken and rotational symmetry emerges. The resulting patterns lack strict periodicity but exhibit long-range order with quasicrystal characteristics, and are therefore termed moiré quasicrystal patterns. In the nonlinear stage, continuous energy accumulation eventually exceeds the threshold required to sustain rotational symmetry, leading to disordered pattern configurations. In our numerics, the energy of ground state is $4.8E_r$, compared to an energy threshold of approximately $7.8E_r$ for pattern stability.

Chin\textit{ et al.} experimentally reported the hexagonal lattice pattern formation in driven BECs, characterized by six distinct modes in momentum space \cite{zhang2020pattern}. In this letter, we investigate the system using the coupled GP equations, where the DW distribution in momentum space condenses to a point. Therefore, we amplify the momentum-space DW via a logarithmic function $\ln{n(\boldsymbol{k},t)}$. This amplified DW distribution similarly exhibits six distinct modes under lattice symmetry, with the modes equally spaced by $\pi/3$ in directions. In particular, each mode comprises two sub-modes, which is a direct consequence of the bilayer lattices. The angle between these sub-modes precisely matches the twist angle of the bilayer lattices (Fig. \ref{density wave amplitude}\textcolor{blue}{(b)} bottom left).

The six modes remain distinct during the preparation stage. Upon entering the excitation stage, these modes progressively develop interconnections. This transformation culminates in the pattern-forming stage where the six modes connect into a unified and discernible hexagram, indicating lattice symmetry breaking and the emergence of a new symmetry in real space. Finally, this configuration becomes largely indistinguishable in the nonlinear stage (Fig. \ref{density wave amplitude}\textcolor{blue}{(b)} bottom row). Although the modulation parameters ($a_m$, $\omega$) supporting the moiré quasicrystal patterns vary, all six modes undergo this transformation. Only the duration of each stage differs. The changes in the patterns in both spaces provide clear validation of these four distinct dynamical stages.

\begin{figure*}
		\centering
		\includegraphics[width=1.5\columnwidth]{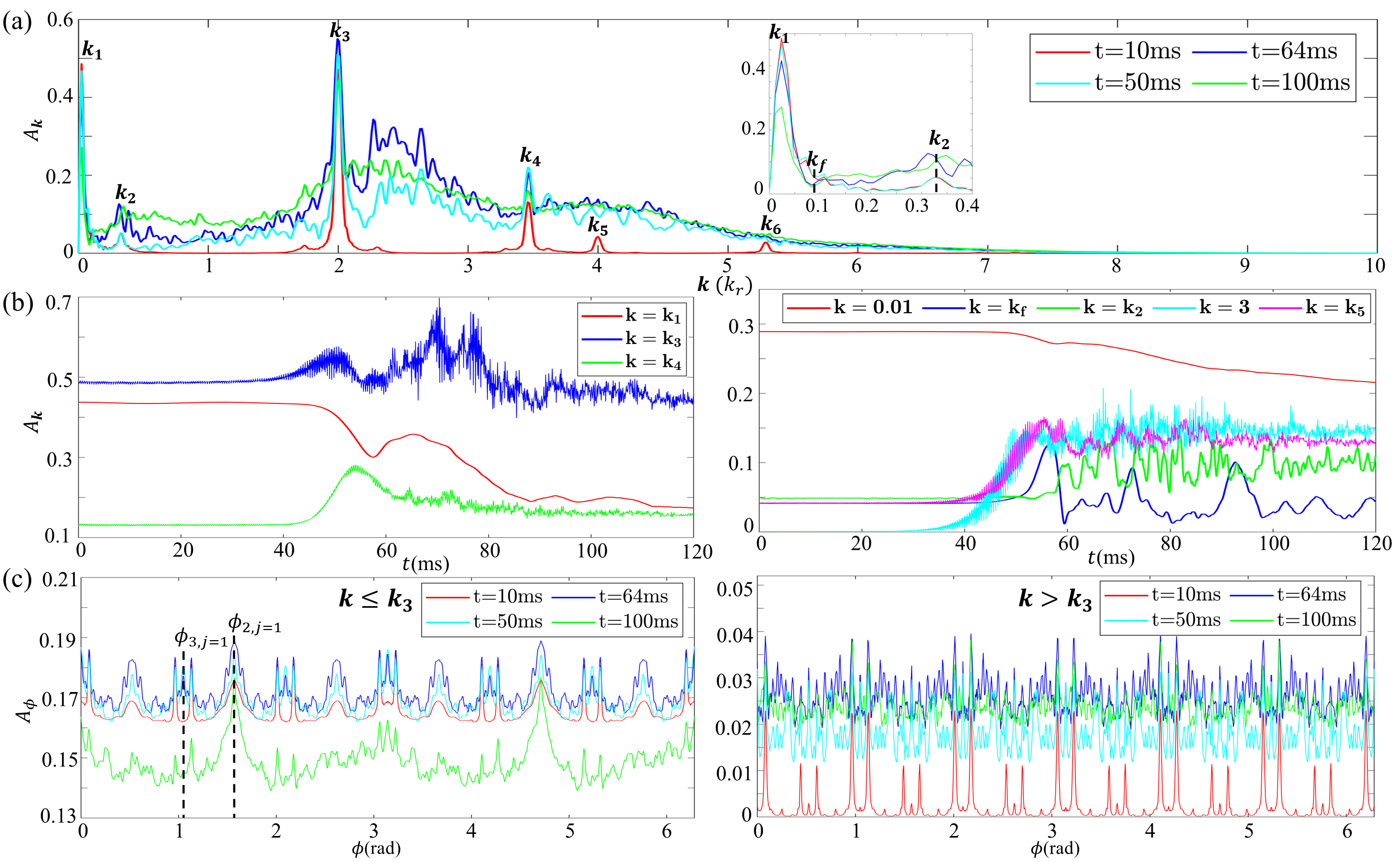}
		\caption{\fontfamily{ptm}\selectfont The radial and angular DW amplitudes with $\omega=200$ Hz and $a_m=50a_0$. (a) The dependence of $A_{\mathbf{k}}$ on $\mathbf{k}$ at $t=10, 50, 64, 100$ ms, where $\mathbf{k}_f=\sqrt{m\omega/\hbar}$ is the the parametric resonance condition which only considers weak atomic interaction term. (b) The dynamical $A_{\mathbf{k}}$ with different momenta $\mathbf{k}$. (c) The dependence of $A_{\phi}$ on $\phi$ when $\mathbf{k}\le\mathbf{k}_3$ (left) and $\mathbf{k}>\mathbf{k}_3$ (right) at $t=10, 50, 64, 100$ ms.}
		\label{akaphi}
	\end{figure*}

In our calculations, the strength of the drive for one layer is $E_g=\left\langle|\psi_1|^2\right|g_{11}(t=0)\left||\psi_1|^2 \right \rangle\approx0.02$, which is weak compared with other energy scales ($E_{V_1}=\left\langle|\psi_1|^2\right|V_1\left||\psi_1|^2 \right \rangle\approx2.36$,  $E_{\Omega}=\left\langle|\psi_1|^2\right|\Omega\left||\psi_1|^2 \right \rangle\approx0.50$). Obviously, the energies keep $E_{V_1}>E_{\Omega}\gg E_{g}$ in the whole dynamical evolution. Therefore, we think the formation of Moiré quasicrystal patterns is associated with momenta of collective modes that are resonantly excited by the periodic drive. To quantitatively analyze these momenta, we  extract the DW amplitude, $A_{\mathbf{k}}=n_0^{-1}\int _{\left | \mathbf{k} \right | }\tilde{n}(\mathbf{k})d \mathbf{k}$ \cite{am}. However, different from the Ref. \cite{am}, our system is more complex and not easy to get an analytical derivation of the  parametric resonance condition from the Mathieu equation. Therefore, we numerically calculate the dependence of $A_{\mathbf{k}}$ on $\mathbf{k}$ during the dynamical evolution. 

For ground state, the DW is controlled by six distinct momenta (Fig. \ref{akaphi}\textcolor{blue}{a} red curve), where the two main momenta ($\mathbf{k}_1=0.03k_r$, $\mathbf{k}_2=2k_r$) correspond precisely to the central mode and its surrounding six modes in the momentum space (Fig. \ref{density wave amplitude}\textcolor{blue}{(b)} bottom line). A multitude of additional collective modes emerge during the pattern-formation stage ((Fig. \ref{akaphi}\textcolor{blue}{a} blue curve). We track the evolution of these modes throughout the dynamics and uncover several interesting regularities.

In the region of $0<\mathbf{k}<\mathbf{k}_1$, $A_{\mathbf{k}}$ keeps decreasing after excitation, indicating these momentum-associated collective modes play a waning influence on pattern formation. In the region of $\mathbf k_1<\mathbf{k}<\mathbf{k}_2$, $A_{\mathbf{k}}$ of original modes exhibits damped-harmonic envelope oscillations around its initial value. In the region of $\mathbf{k}>\mathbf{k}_2$ (except for $\mathbf{k}=\mathbf{k}_3,\mathbf{k}_4$), the trend of $A_{\mathbf{k}}$ is similar to that of $A_{\mathbf{total}}$, indicating that collective modes with these momenta are dominant in the excitation (Fig. \ref{akaphi}\textcolor{blue}{b} right). Normally, in the absence of mode coupling, $A_{\mathbf{k}}$ is expected to return to near its initial value or exhibit a pronounced drop after entering the nonlinear stage. Thus we think these new collective modes are coupled each other and eventually cannot be decoupled. In the above region, the weak and peakless oscillations of $A_{\mathbf{k}}$ also reflect radial coupling among the modes. Notably, the three modes ($\mathbf{k}_1,\mathbf{k}_3,\mathbf{k}_4$) of the ground state retain distinct peaks throughout the whole dynamical evolution. Obviously these three modes have no radial coupling to other additional collective modes, leading to the decline in $A_{\mathbf{k}_{1,3,4}}$ because of the energy transfer from these modes to the remaining uncoupled modes (Fig. \ref{akaphi}\textcolor{blue}{b} left).

From Fig. \ref{density wave amplitude}\textcolor{blue}{(b)}, we observe the new modes emerge along specific directions when $\mathbf{k}\le \mathbf{k}_3$. Inspired by the dynamical patterns in momentum-space, we wonder whether the newly appearing collective modes are coupled angularly. To quantify the distribution of these modes along the angular direction, we also evaluate $A_{\phi}=\int_\phi \tilde{n}'(\mathbf{\mathbf{k}_\phi}) d\mathbf{\mathbf{k}_\phi}$, where $\tilde{n}'(\mathbf{\mathbf{\mathbf{k}_\phi}})=\int n(\mathbf{r})e^{-i\mathbf{k}_\phi\cdot \mathbf{r}} d\mathbf{r}/2\pi$ with magnitude $\left|\mathbf{k}_\phi\right|=\left|\mathbf{k}\right|$ and angle $\phi$. Consistent with our expectations, new modes emerge along the directions ($\phi_2,\phi_3$) defined by $\mathbf k_2$ and $\mathbf k_3$ (Fig. \ref{akaphi}\textcolor{blue}{c} left), where $\phi_2=\frac{(2j+1)\pi}{6}$ and $\phi_3=\frac{j\pi}{3}$ with $j=0,1,...,5$. Interestingly, for $\mathbf{k}>\mathbf{k}_3$, $A_{\phi}$ consists of a series of distinct peaks (Fig. \ref{akaphi}\textcolor{blue}{c} right), which means the collective modes are angularly uncoupled.

\subsection{The real-space Moiré quasicrystal patterns}
\begin{figure}
		\centering
		\includegraphics[width=1\columnwidth]{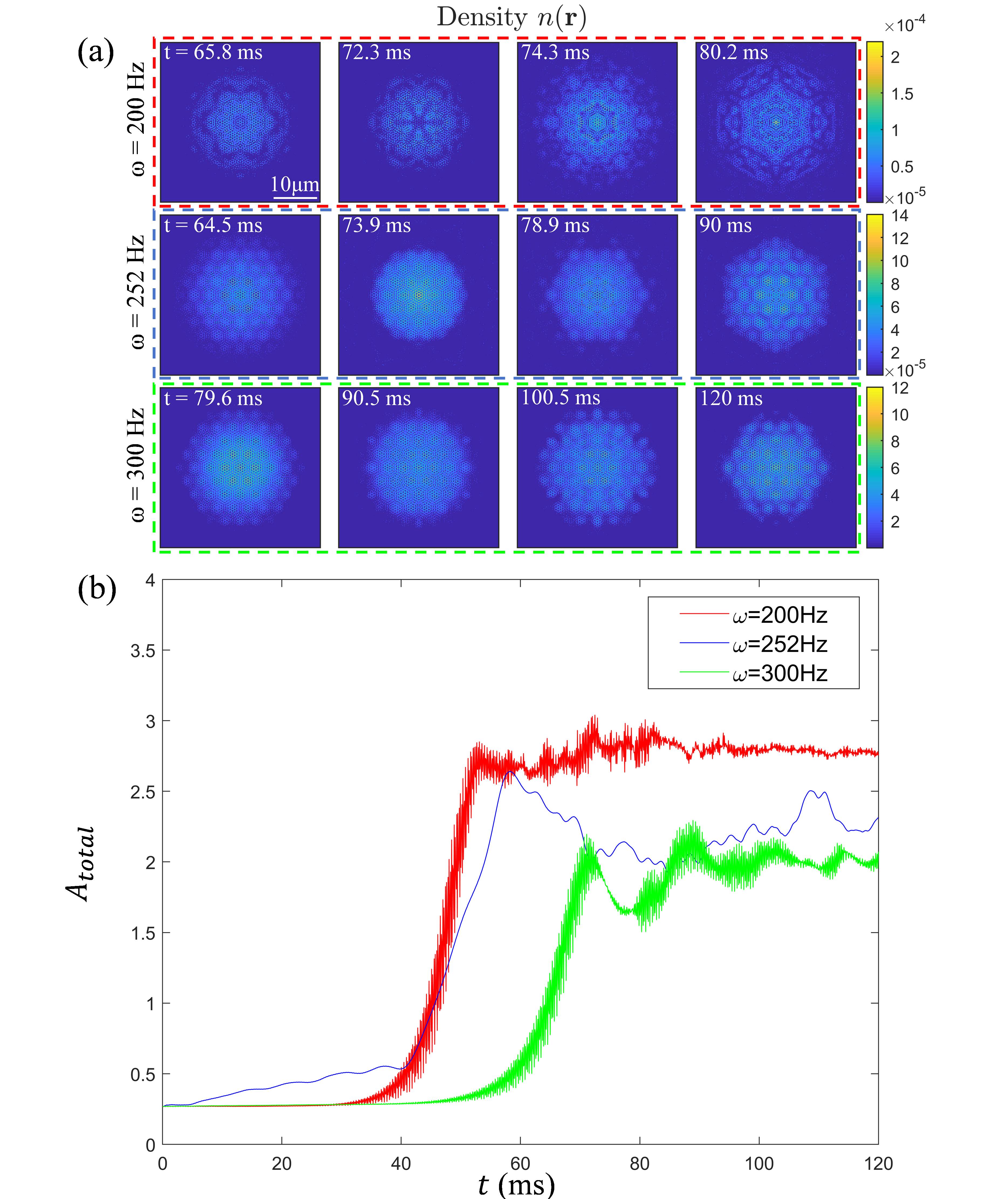}
		\caption{\fontfamily{ptm}\selectfont The dynamical evolution of real-space patterns with different modulation frequencies. (a) The patterns with $a_m=50a_0$ and $\omega=200$ Hz (top), $\omega=252$ Hz (middle) and $\omega=300$ Hz (bottom). (b) The corresponding $A_{total}$} of (a).
		\label{frequency}
	\end{figure}

The rotational symmetry of the patterns significantly depends on the modulation amplitude $a_m$ and frequency $\omega$. We first fix the modulation amplitude and vary the frequency. Increasing the modulation frequency breaks the rotational symmetry of the system. For example, at $a_m=50a_0$ and low $\omega$, the pattern exhibits $D_6$ symmetries and other distinct $D_6$ patterns subsequently emerge during the third stage. Each of these non-subharmonic patterns represent a new symmetrical structure. Upon increasing $\omega$ to approximately 252 Hz, the $A_{total}$ decreases significantly, indicating only a minor DW alteration and patterns with lattice characteristics (Fig. \ref{frequency}\textcolor{blue}{(b)} blue curve). The blue curve in Fig. \ref{frequency}\textcolor{blue}{(b)} is notably smoother than the other two. We think this frequency matches exactly with the inherent frequency of a collective excitation mode. Under this frequency, system is excited without a preparation stage. In other words, energy is injected into the system directly and efficiently from the drive field, causing a smoother curve. This phenomenon differs from the collective mode momentum of DW yet does not conflict with it. For frequencies above this frequency, the formation of moiré quasicrystal patterns is suppressed, while the duration of the lattice patterns is extended.

\begin{figure}
		\centering
		\includegraphics[width=1\columnwidth]{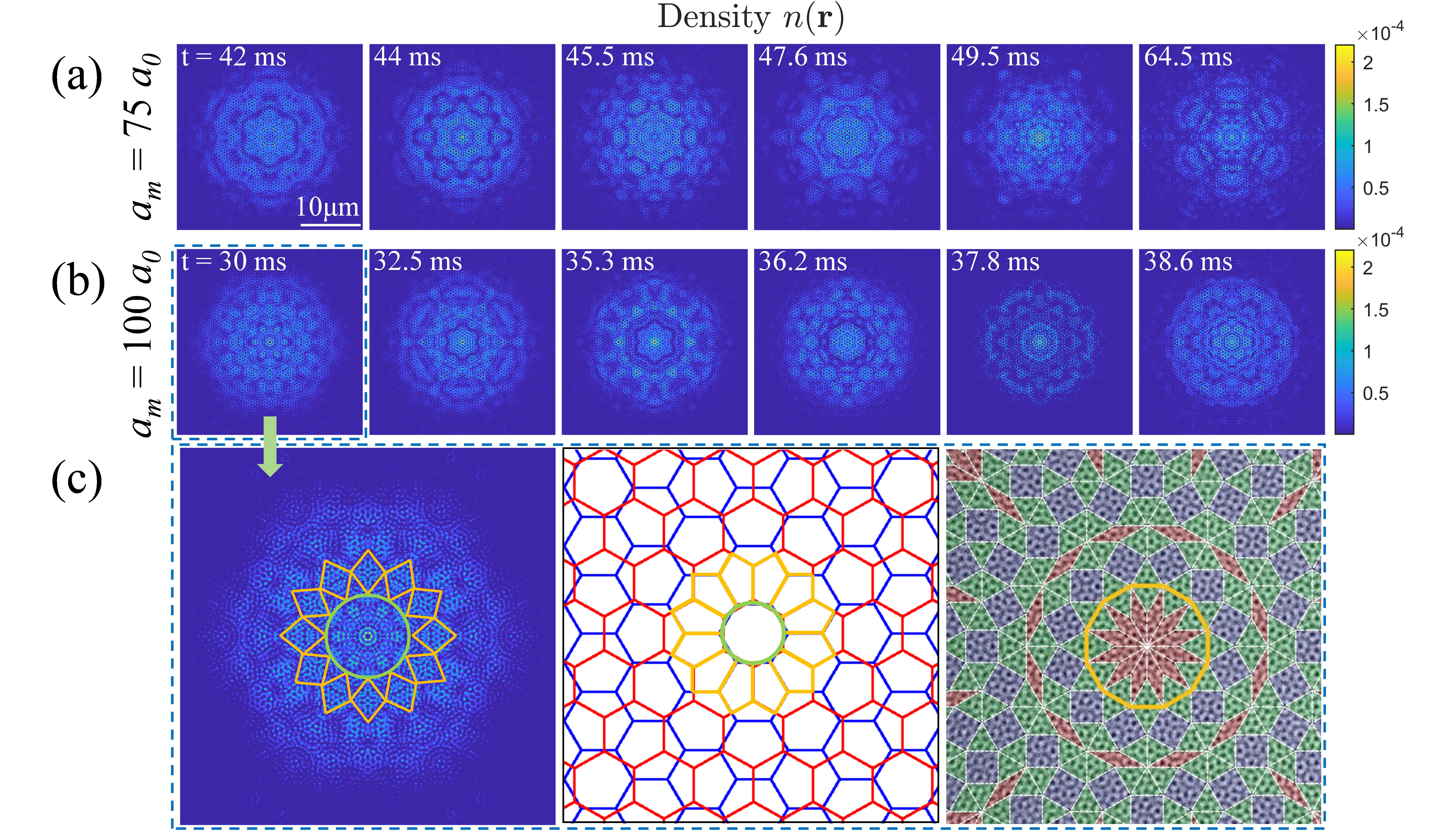}
		\caption{\fontfamily{ptm}\selectfont The dynamical evolution of real-space patterns with different modulation amplitudes. The patterns with $\omega=200$ Hz and (a) $a_m=75a_0$, (b) $a_m=100a_0$. (c) $D_{12}$ moiré quasicrystal patterns comparison: simulations (left), twisted bilayer graphene atomic structure (middle) and graphene quasicrystal (right) \cite{graphenequasicrystal1}.}
		\label{amplitude}
	\end{figure}

We then fix the modulation frequency at $\omega=200$ Hz and vary the modulation amplitude. Increasing $a_m$ yields patterns with increasingly complex symmetry structures. As previously demonstrated, a simple $D_6$ pattern forms at $a_m=50a_0$, which means regions other than the center possess only a single layer of $D_6$ symmetries. At $a_m=75a_0$, two additional layers of $D_6$ patterns develop. These patterns are not simple six-petal types but rather a more complex cobweb-like structures (Fig. \ref{amplitude}\textcolor{blue}{(a)}). 

Remarkably, the system forms a $D_{12}$ pattern at 30 ms with $a_m=100a_0$ (Fig. \ref{amplitude}\textcolor{blue}{(b)}), closely resembling the moiré quasicrystal realized in twisted bilayer graphene rotated by exactly $30^{\circ}$. This pattern exhibits three distinct layers, whose structure is nearly identical to graphene quasicrystal. In twisted bilayer graphene, the stability of hexagonal atomic structure yields standard Stampfli quasicrystal constructions \cite{stampfli,stampfli2}, including triangles, rhombuses, and squares. However, in our driven system, the excited DW presents apparent fluctuations, preventing the formation of patterns with such regular shapes. For a clearer comparison with graphene quasicrystal, we outline the contours of the second layer in Fig. \ref{amplitude}\textcolor{blue}{(c)}. The patterns retain $D_6$ symmetries at other times.

In general, increasing the modulation amplitude while appropriately decreasing the frequency promotes lattice symmetry breaking and the emergence of rotational symmetry. This process also increases the multiplicity of rotational symmetry, yielding more complex symmetrical structures. For instance, the pattern maintains its lattice symmetry at $a_m=50a_0$ and $\omega=252$ Hz. Further increasing the modulation amplitude leads to the emergence of quasicrystal patterns. The stable parameter region for moiré quasicrystal patterns agrees well with the regularity of Floquet stable tongues derived from the Mathieu equation (black dotted line in Fig. \ref{mathieu}) \cite{tongue1,tongue2}. The parameters (blue and green lines in Fig. \ref{frequency}\textcolor{blue}{(b)}) are beyond this stable region, with no moiré quasicrystal patterns emerging.

\subsection{The momentum-space Moiré quasicrystal patterns}
\begin{figure}
		\centering
		\includegraphics[width=1\columnwidth]{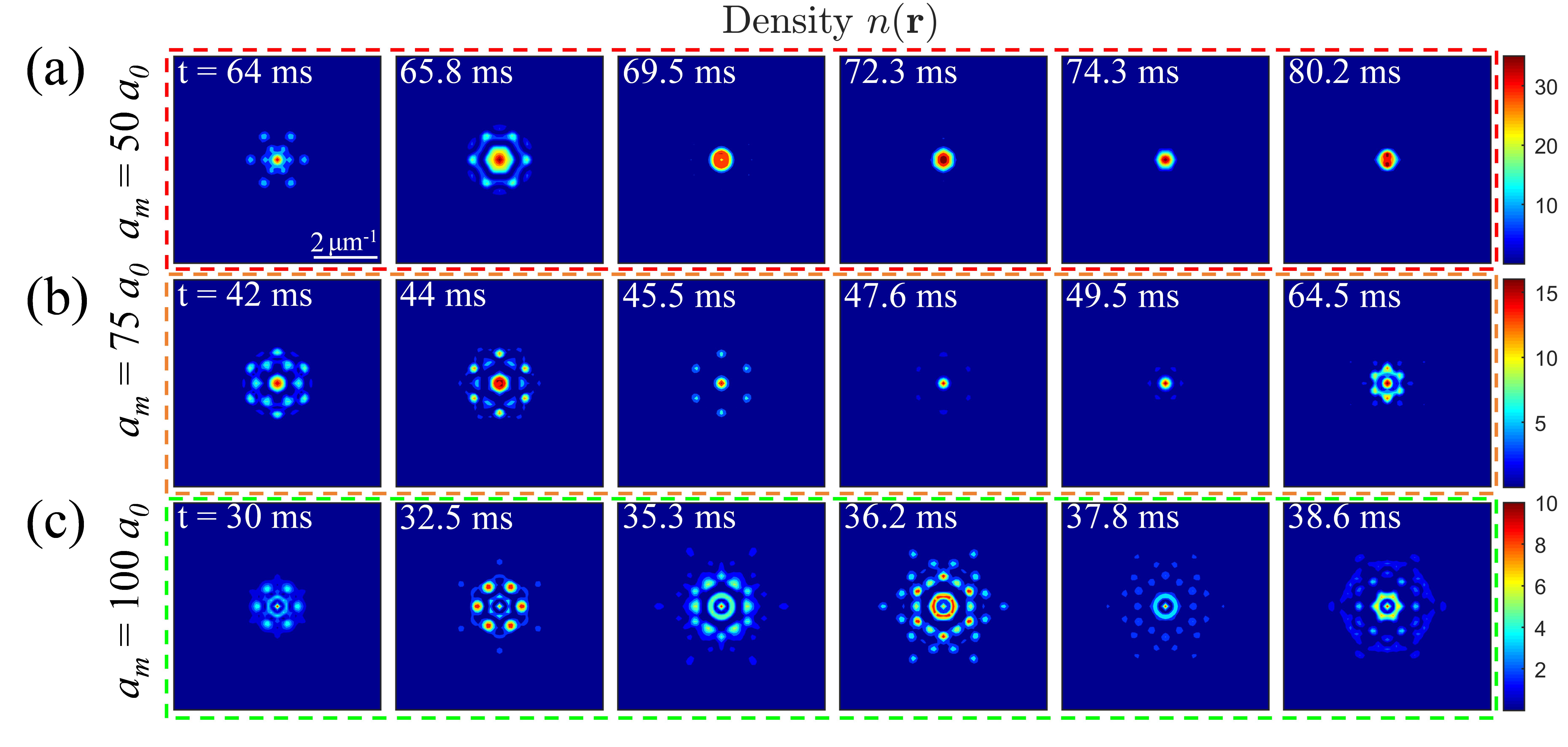}
		\caption{\fontfamily{ptm}\selectfont The dynamical evolution of momentum-space patterns with different modulation amplitudes. The patterns with $\omega=200$ Hz and (a) $a_m=50a_0$, (b) $a_m=75a_0$ and (c) $a_m=100a_0$.}
		\label{momentum}
	\end{figure}

\begin{figure}[t!]
		\centering
		\includegraphics[width=1\columnwidth]{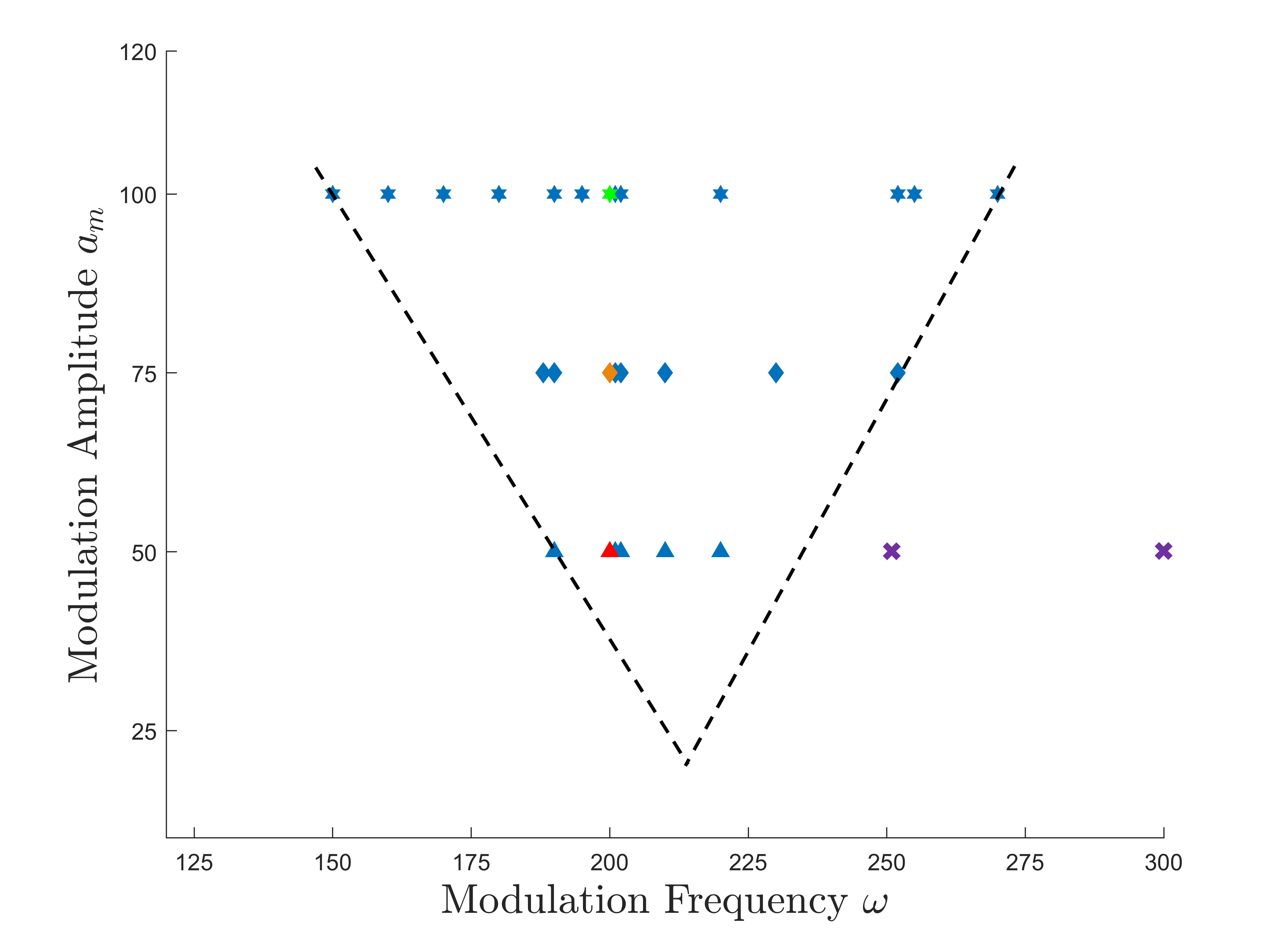}
		\caption{\fontfamily{ptm}\selectfont  The stable parameter region for moiré quasicrystal patterns is shown, with numerical results in blue shapes and the boundary marked by black dotted lines. The patterns displayed in Fig. \ref{momentum} correspond to $\omega=200$ Hz and $a_m=50a_0$ (red triangle), $a_m=75a_0$ (brown rhombus) and $a_m=100a_0$ (green hexagram). The purple crosses outside the stable region correspond to $a_m=50a_0$ and $\omega=252$ Hz, $\omega=300$ Hz in Fig. \ref{frequency}\textcolor{blue}{(b)}.}
		\label{mathieu}
	\end{figure}

We further analyze the symmetry of the DW distribution in momentum space based on Fourier analysis $\psi(\boldsymbol{k})=\int e^{-i\mathbf{k}\cdot \mathbf{r}}{\psi}(\mathbf{r})d\mathbf{r}/2\pi$. The Fourier DW $n(\boldsymbol{k})$ also exhibits high rotational symmetry, typically forming $D_6$ patterns. These momentum-space patterns are consistent with real-space patterns at specific times. At $a_m=100a_0$ and $\omega=200$ Hz, the momentum-space pattern forms a hexagram nearly identical to the real-space pattern at 36.2 ms, but rotated clockwise by $60^\circ$. Similarly, at $a_m=75a_0$ and $\omega=200$ Hz, the patterns in both spaces match well at 42, 44, and 64.5 ms (Fig. \ref{momentum}\textcolor{blue}{(b)}). At 64.5 ms in particular, both patterns exhibit a symmetrical butterfly structure. The momentum-space patterns also do not exhibit subharmonicity. However, unlike the real-space patterns, some momentum-space patterns exhibit only $D_4$ symmetries at the center.

\section{Conclusions and Discussions}
We numerically study the formation of novel moiré quasicrystal patterns of driven ultracold Bose gasses in twisted bilayer hexagonal lattices. Dynamical evolutions of the MSD and total DW amplitude reveal the stage for the emergence of moiré quasicrystal patterns. In our weakly driven system, pattern formation is closely related to the momenta of collective modes. By quantifying the radial and angular DW amplitudes, we obtain the distributions of the emerging collective modes. We conclude that these new collective modes are coupled only radially and eventually cannot be decoupled. The system undergoes a symmetry transition from lattice symmetry breaking to rotational symmetry emergence during the modulation, ultimately forming novel moiré quasicrystal patterns. Each moiré quasicrystal pattern generally persists for less than 1 ms before transforming into the next symmetrical configuration. This symmetry transition heralds a new quantum dynamical phase transition.

The spatial symmetry of patterns evolves temporally when the frequencies or amplitudes are modulated. Increasing the modulation frequency prevents the formation of moiré quasicrystal patterns while preserves the original lattice characteristics. In contrast, a larger modulation amplitude generates more complex symmetrical structures. In particular, a $D_{12}$ pattern emerges at $a_m=100a_0$ and $\omega=200$ Hz, closely resembling the moiré quasicrystal in twisted bilayer graphene. Furthermore, the momentum-space patterns also exhibit high rotational symmetries, consistent with the real-space patterns at specific evolution times.

In summary, our results demonstrate that Floquet-engineered intralayer atomic interactions can generate novel quasicrystal patterns in bosonic systems without the need for additional quasiperiodic potentials. This phenomenon arises from the competition between the drive-induced disorder and the superfluidity of the system. The dynamical evolution of quasicrystal patterns provides a unique perspective for investigating quantum dynamical phase transitions in quasicrystalline systems. Furthermore, our numerical results reveal a sensitivity of the pattern formation to the modulation frequency, where a variation of only 1 Hz is sufficient to produce completely distinct moiré quasicrystal patterns. This work provides a new quantum platform for exploring quasicrystals and their symmetry characteristics in ultracold bosonic systems. The anyonic behaviors supported by the quasicrystals could offer a promising avenue for topological quantum computation.

\textit{Acknowledgments.} Our work is supported by the Natural Science Foundation of Shanghai (Grant No. 23ZR1418700), the National Key Research and Development Program of China (Grant No. 2025YFF0515201), the Joint Funds of the National Natural Science Foundation of China (Grant No. U25D8014), the National Natural Science Foundation of China (Grant No. 11774093), the Program of Chongqing Natural Science Foundation (Grant No. CSTB2022NSCQ-MSX0585), and the Innovation Program of Shanghai Municipal Education Commission (Grant No. 202101070008E00099).


\bibliography{ref}

\end{document}